# Imaging High Jitter, Very Fast Phenomena: A Remedy for Shutter Lag

Noah Hoppis,[1] Kathryn M. Sturge,[1] Jonathan E. Barney,[2] Brian L. Beaudoin,[1] Ariana M. Bussio,[1] Ashley E. Hammell,[1] Samuel L. Henderson,[2] James E. Krutzler,[1] Joseph P. Lichthardt,[2] Alexander H. Mueller,[2] Karl Smith,[2] Bryce C. Tappan,[2] and Timothy W. Koeth[1,2,a]

[1]*Department of Materials Science and Engineering, University of Maryland, College Park, Maryland, 20742, USA*

[2]*Los Alamos National Laboratory, Bikini Atoll Rd, Los Alamos, New Mexico, 87545, USA*

Dielectric breakdown is an example of a natural phenomenon that occurs on very short time scales, making it incredibly difficult to capture optical images of the process. Event initiation jitter is one of the primary challenges, as even a microsecond of jitter time can cause the imaging attempt to fail. Initial attempts to capture images of dielectric breakdown with a gigahertz frame rate camera and an exploding bridge wire initiation were stymied by high initiation jitter. Subsequently, a novel optical delay line apparatus was developed in order to effectively circumvent the jitter and reliably image dielectric breakdown. The design and performance of the optical delay line apparatus are presented. The optical delay line increased the image capture success rate from 25% to 94% while also permitting enhanced temporal resolution and has applications for use in imaging other high-jitter, extremely fast phenomena.

_______________________________

[a] Author to whom correspondence should be addressed. Electronic mail: koeth@umd.edu.

## I. INTRODUCTION

Imaging very fast phenomena with nanosecond time resolution is challenging for several reasons—the events are incredibly brief such that their evolution cannot be captured with a typical camera, and many phenomena do not have practical or predictable triggering mechanisms. One such phenomenon that has proven difficult to image is mechanically-initiated dielectric breakdown. Dielectric breakdown is the process by which an insulating material rapidly transitions to a highly conductive state when subjected to electric potential gradients above the material-dependent dielectric breakdown threshold (Knaur and Budenstein 1980). The current passing through the material during the breakdown generates both optical emissions and permanent tree-like damage patterns (Gross 1958, Furuta, Hiraoka and Okamoto 1966, Akishin, et al. 1977, Noskov, et al. 2002). Figure 1 shows the channel formations in a dielectric sample after breakdown. The optical emissions—seen as a brilliant flash of light to the naked eye—occurs in the same location as the damage pattern (Cooke, Space-charge-induced Breakdown in Dielectrics 1986). Thus, imaging of the visible light emissions as breakdown occurs allows for the measurement of breakdown pattern formation.

Initial attempts were made to capture images of a dielectric breakdown event (DBE), and the corresponding formation of damage patterns with a high-speed framing camera at Los Alamos National Laboratory. The breakdown process was initiated in electron-loaded polymethyl methacrylate (PMMA) by mechanical shock from the explosion of an exploding bridgewire detonator (EBW). The EBW was chosen for its precise timing and electrical firing control, which would permit camera synchronization. However, it was discovered that there exists substantial variation in time, around 0.7 µs, between the EBW detonation and the initiation of breakdown in the dielectric. This breakdown initiation jitter time was greater than the duration of the DBE itself. As a result, the majority of images did not capture the desired breakdown dynamics; the framing camera either began taking photos too early or too late.

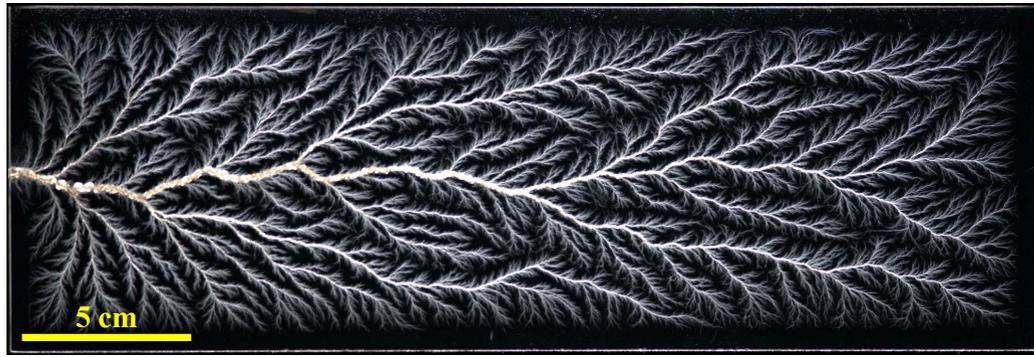

Figure 1. Discharge pattern formed in a PMMA sheet 30.5 cm long by 10.2 cm high by 2.5 cm thick; the same size used in this experiment. The sample is lit from the left, behind the discharge initiation point, to illuminate the dendritic pattern formed by the dielectric breakdown event (DBE).

To circumvent this jitter time, an Optical Delay Line (ODL) was developed and introduced. Previous work used ODLs as a method for time resolving (Lorenz, Dymoke-Bradshaw and Dangor 1997), or simultaneous interrogation (Woodworth, et al. 2004). This work extends the ODL to compensate for the timing limitations of imaging systems. The discharge was initiated via pneumatic punch and the breakdown current from the discharge was measured via a fast current transformer. The current signal was a reliable timing indicator for imaging and was used to trigger the framing camera, which has a minimum 65 ns shutter lag between trigger and first image capture. To account for this extra time, a plane mirror was positioned opposite the camera and sample, so that the light from the discharge traversed half the requisite number of light seconds to the mirror and was then reflected into the camera at the end of the shutter lag period.

## II. METHODS AND MATERIALS

The samples imaged in this work were rectangular prisms of commercially procured PMMA (Pease Plastics) with dimensions of 305 mm by 102 mm for the imaged front face and thickness 25 mm with a tolerance of ±3 mm on all lengths.

The samples were irradiated via electron beam using the 4.5 MeV electron Dynamitron at E-Beam Services in Lebanon, Ohio. The electron beam charged each sample in the direction normal to the 305 mm by 102 mm face which was to be imaged. Each sample was individually loaded with charge densities ranging between 10 and 18 mC/m$^2$ at the midplane. Once loaded and transported to Los Alamos National Laboratory, dielectric breakdown of these charged samples was initiated by mechanical insult. This insult resulted in the expulsion of the retained charge from the plastic through the region of mechanical damage, as the dielectric breakdown figure grew outward into the remaining undamaged material.

In order to capture images of the extremely rapid DBE, state-of-the-art framing cameras are required. A Specialised Imaging SIMX16 framing camera capable of 3 ns minimum exposure for 16 beam splitter coupled CCDs (Specialised Imaging 2023) was used for this work. One CCD in the camera was disabled due to prior instrument damage, so only 15 individual image frames were captured for each discharge.

The image-intensified CCDs in the SIMX16 camera must receive a trigger signal at least 65 ns prior to beginning exposure. This is known as the "shutter lag". The shutter lag could be increased beyond 65 ns with the camera software, but it could not be reduced below 65 ns. This delay presents the central challenge to imaging dielectric breakdown. Non-zero shutter lag requires precommitment to a designated trigger time. But if the timescale of the event being captured is short compared to the event initiation jitter time, most attempts to image the event will capture images from either before or after the event, rather than of the event itself. Thus, there are two options: (if possible) minimize the event jitter by modifying the initiation process or develop a method to coordinate the initiation time to the camera shutter time on a shot-by-shot basis.

### A. Exploding Bridgewire Triggering of Dielectric Breakdown Events

The first approach taken to image a DBE was to use an EBW. The system used trigger breakdown and image the resultant event is shown in figure 2. EBWs have short risetimes with low jitter, release hundreds of joules of energy on detonation, and are commercially available (Smilositz, et al. 2020). The EBWs were coupled to the side face of the sample by a fixture such that the emitted shockwave would enter the sample normal to the center of the 102 mm by 25 mm face. The RISI Teledyne RP-2, with a specified 1.65 µs time to shockwave emission with a standard deviation of 35 ns (Jenkins, et al. 2013) was chosen as the primary detonator used in these experiments. In order to reliably record the DBE, the SIMX16 shutter lag was increased to match the EBW function time.

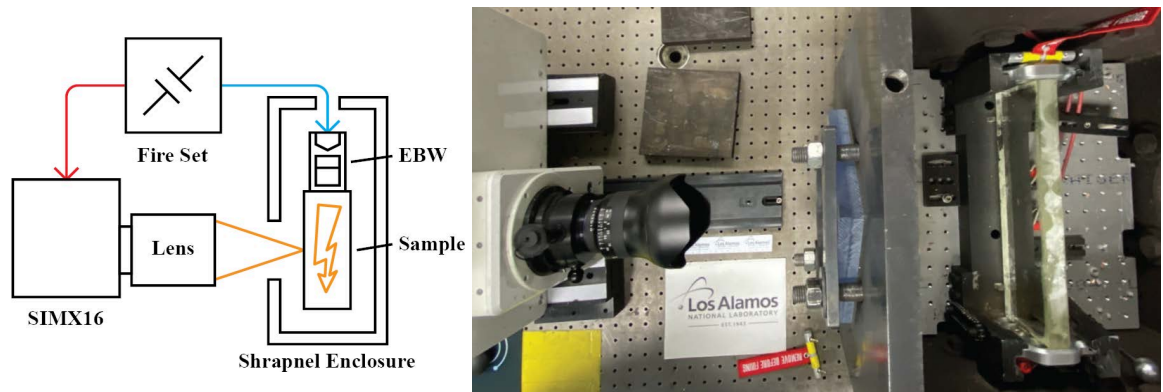

Figure. 2. Left: a wiring diagram of the EBW discharge initiation fixture; Right: the system as implemented.

In total, 84 DBEs were initiated using the EBW system, of which 1260 images were captured (15 for each discharge). Of these images, only approximately 25% contained breakdown evolution; many were dark images before breakdown occurred, or images of already formed plasma channels after the breakdown. Analysis of the images and chosen camera shutter lags revealed that the breakdown initiation jitter time—the standard deviation of times between EBW detonation and first light released by the event—is approximately 0.7 µs. Due to this jitter, photographing the DBE on any of the 15 frames when recorded over a time span less than 500 ns was difficult. Because breakdown travels longitudinally through the dielectric bulk (Cooke, Williams and Wright, Electrical Discharge Propagation in Space-Charged PMMA 1982) this record length limitation also imposes a limit on the spatial resolution which may be recorded given the finite resolution of the camera CCDs. Based on these results it was decided that the EBW system as implemented was not sufficient for repeatable imaging of the DBEs.

Due to the initiation jitter time limitations of the existing EBW discharge system, subsequent work was focused on developing a method to coordinate the opening of the shutter and the beginning of the event on a shot-by-shot basis, carefully circumventing the issue of initiation jitter. To accomplish this, the ODL was created.

**B. Optical Delay Line Apparatus**

The ODL apparatus involves the use of a plane mirror and reflected light from the sample in order to coordinate the timing of the DBE with the camera shutter lag. A schematic and photograph of the ODL system are shown in figure 3. In these experiments, the DBE was initiated in a fixture with the framing camera located on the same bench. Instead of using the camera to image the sample directly, the framing camera was instead focused on the sample by way of plane mirror placed on another bench at a predetermined distance. The exact time of breakdown initiation is observed by a sensor and this sensor triggered the camera shutter. In this experiment, the output current from the DBE was measured and used to trigger the camera, but other triggering mechanisms are possible (e.g. a fast photodiode detecting first breakdown light). The light from the DBE traveled to the plane mirror before being reflected back into the camera lens by way of a telescope, which allowed for the desired millimeter-level spatial resolution in the final images. This additional distance traveled by the light delayed the optical signal from the sample from reaching the camera by a few hundred nanoseconds; just enough time to allow the camera to prepare for imaging. The distance of the optical path and the shutter lag of the camera were adjusted such that the light reached the camera at the exact moment the full shutter lag time had elapsed.

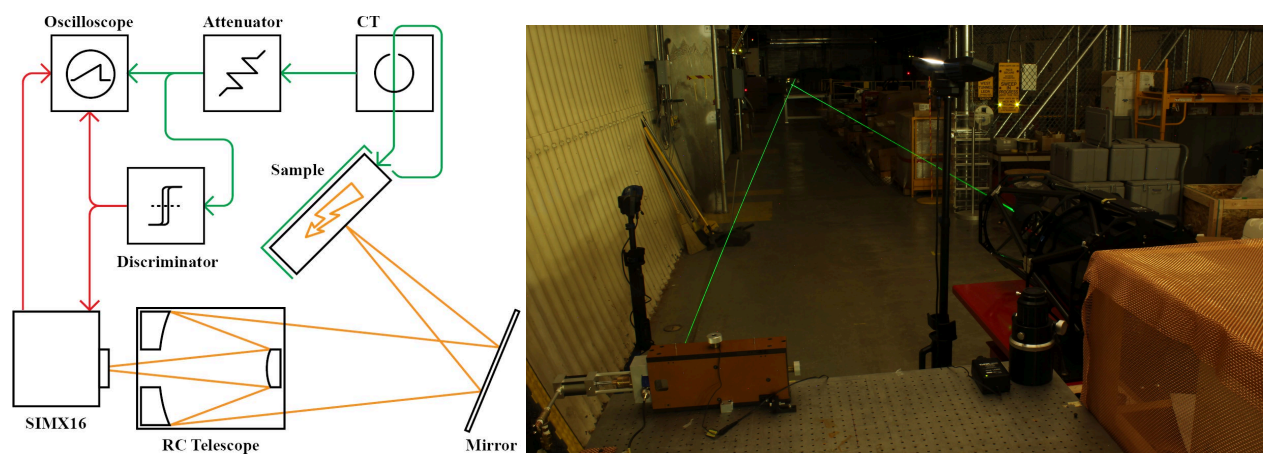

Figure. 3. Left: a schematic of the ODL system showing the triggering system, right: the ODL system as implemented, with a green laser showing the optical path between the fixture (bottom left), the mirror (top), and telescope (middle right).

The camera trigger signal was derived from the secondary signal of a current transformer (CT) (Com-Power CLCE-1032), which measured the DBE current. To capture the majority of discharge current, all faces except the single 305 mm by 102 mm imaged front face were cradled by a close-fitting aluminum enclosure which minimized discharge inductance. The mechanical impact to initiate breakdown was provided by a pneumatic cylinder which struck a brass rod with a pointed tungsten tip. The body of the rod passed through the center of the CT, and the tip of the rod was driven approximately 2 mm into the center of a 102 mm by 25 mm face. The rod was connected by a short phosphor bronze bellows to the rest of the metal enclosure by a path around the outside of the CT, forming a single, low inductance primary turn.

The measured current signal from the CT secondary was attenuated and split with a 6 dB power splitter (Midwest Microwave 2533). Half of the signal was recorded by the experiment oscilloscope (Keysight MSOX6004A) to capture the current traces for each trial, while the other half traveled to a fast linear discriminator (Ortec 436) which generated a TTL pulse to trigger the SIMX16. When image acquisition begins, the SIMX16 also outputs a TTL signal with a width corresponding to the image record length which was also recorded by the oscilloscope in order to verify shot timing. The relative timing of these timing signals as recorded by the experimental oscilloscope is shown in figure 4.

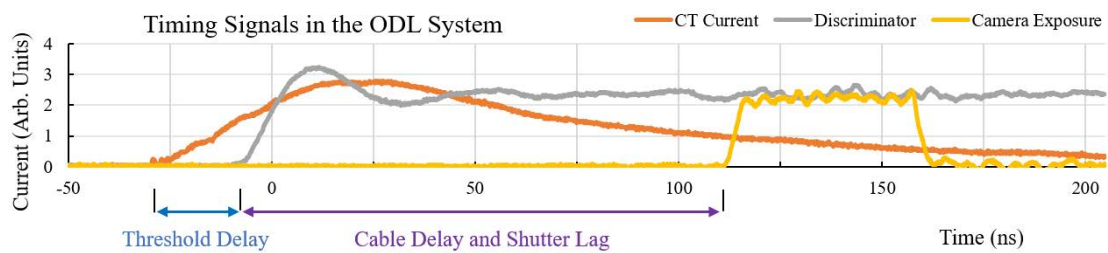

Figure 4. A typical oscilloscope recording of the breakdown current (orange) and discriminator (grey) and camera (yellow) timing signals.

To compensate for the pre-trigger time and trigger generation delay, a minimum ODL distance of 35 m is necessary between the mirror and the sample. This results in a round-trip distance of 70 m, giving the necessary delay to accommodate the trigger circuit propagation time and shutter lag. However, in practice, a longer ODL distance of 45 m (90 m round trip) was used as a means to allow the camera to begin recording slightly before the event initiation, with one blank frame produced prior to the first light reaching the camera. It is important for the ODL and system cables to be close to the minimum practical length in order to maximize imaging resolution and CCD exposure. Propagation delays through the system cables were measured with time domain reflectometry using the Keysight MSOX6004A oscilloscope, and are shown in Table I.

Table I. Cable propagation delays between trigger components (nanoseconds). "In" refers to the input to the specified device; "out" refers to the output of the specified device.

|  | *Discriminator In* | *SIMX16 In* | *Oscilloscope Channel In* 1 | 2 | 3 |
|---|---|---|---|---|---|
| *Current Transformer Out* | 13 | - | 19 | - | - |
| *Discriminator Out* | - | 6 | - | 7 | - |
| *SIMX16 Out* | - | - | - | - | 18 |

For a distance of 45 m, to achieve the desired 0.5 mm spatial resolution and 10:1 signal-to-noise ratio (SNR), a 406 mm aperture, f/8 Ritchey Chrétien telescope (Guan Sheng Optical 406 mm F/8 RC OTA) was used as the image forming optic for the SIMX16 framing camera. Beyond removing Crayford focuser to permit direct imaging, no modifications to the telescope were required. Due to the finite distance to the sample, it has been calculated that the working f-number of the imaging system was increased to f/8.3, making the usual unity pupil magnification assumption. The theoretical light gathering efficiency of the ODL is close to $1:10^{-6}$, about three orders of magnitude less than the light gathering efficiency of the EBW imaging system. However, for the distances used in the ODL and the particular luminous intensity of the DBE, it was found that aperture-limited imaging performance was bottlenecked by resolution rather than light gathering efficiency, in part due to the image intensifier. Because the object distance was quite close and the telescope was designed for focusing at infinity, the telescope and camera were placed on separate tables to achieve the 0.7 m required back focus distance. Because the telescope and camera were both very large, direct mechanical connection between the two through extension tubes would have frustrated the fine motions required to focus the system. Instead, baffling was placed between the discharge fixture and camera, and the laboratory lights were turned off.

**III. OPTICAL DELAY LINE RESULTS**

In total, the ODL was used to record 116 dielectric breakdown events. A typical image series of a breakdown event is shown in figure 5. Of these, only four of the events failed to trigger the camera. The cause of these four failures remains unknown. Of the 116 shots imaged during this experimental campaign, 94% of acquired images contained visible breakdown evolution. When contrasted with the 25% successful imaging rate with the EBW imaging system previously used in this work, this represents dramatic improvement in system reliability. Additionally, the ODL allowed for the minimum framing camera capture time (time from beginning to end of the photo capture sequence) to be reduced from 500 ns to 17 ns. This is the minimum record length capability of the SIMX16; exposure time is set to 3 ns, and the exposure times are overlapped by 2 ns. Because the total capture time was reduced to 17 ns, one nanosecond temporal resolution was achieved, allowing for greatly enhanced analysis of sequential breakdown behavior.

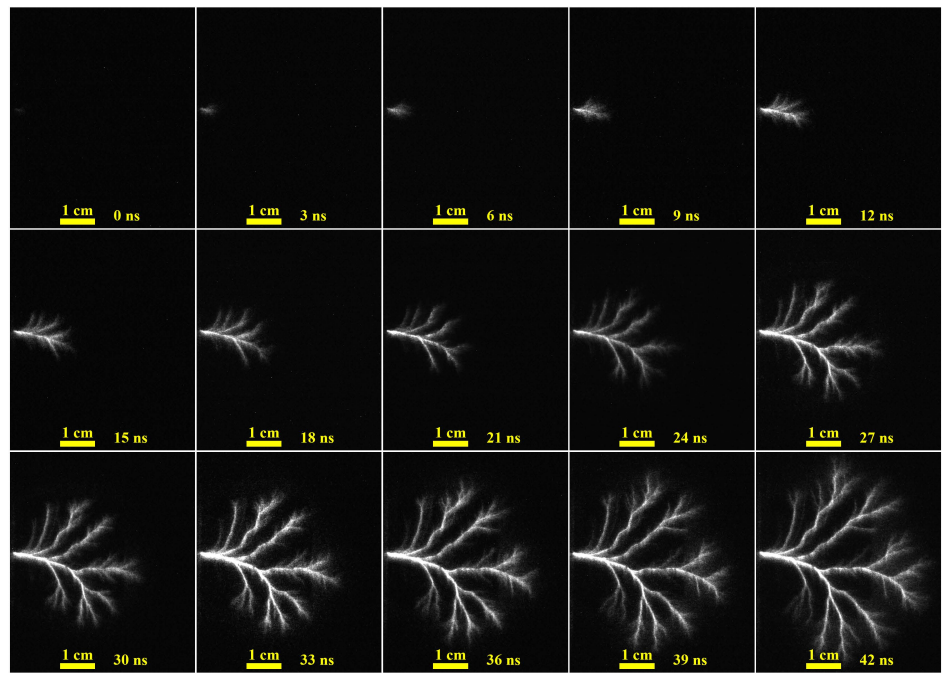

Figure 5. An image sequence captured using the ODL system. Exposure time was 3 ns, 0 ns interframe time. Image sequence is of the sample shown in figure 1.

## IV. REMAINING CHALLENGES AND FUTURE WORK

Several important limitations of the experimental apparatus described above should be noted. Among these are limited spatial resolution, the inability to image dim or non-luminous events, and inherent light gathering limits due to the distance between the source and imaging aperture. It is expected that further improvements will remediate many of these limitations.

The spatial resolution of the ODL system was still substantially worse than the angular resolution of the telescope (the Dawes limit). The reflecting mirror opposite the telescope suffered from distortions (exaggerated by thermally induced stresses due to low temperatures in the laboratory during the scheduled dates for the experiment) which caused poor optical collimation. For several event image captures, some correction of this collimation was achieved by obstructing the areas of the mirror with highest distortion to reach a compromise between CCD illumination (and thus SNR) and resolution. A new mirror with good flatness and thermal characteristics along with a more accommodating mount should both improve the spatial resolution and light gathering ability of the existent system.

As shown in Figure 5, during the first experimental run in which the EBW apparatus was used, some samples were back-illuminated by a 400 W 640 nm laser (Specialised Imaging SI-LUX640). The shadowgraph worked similarly to the focused shadowgraph as described by Settles (Settles 2001), with the point source replaced by the laser and a focusing lens. Because the laser was pulsed, timing the laser illumination to begin before the DBE was not possible, as such it could not be used in conjunction with the ODL method. This exclusion was regrettable because it prevented the ODL apparatus from imaging any components of the ESD which were not strongly self-luminous. Because many other events of interest are not inherently bright, it would be very desirable to circumvent this limitation. Future systems could make use of a continuous laser source, or a secondary trigger source related to the system for mechanically stimulating breakdown (such as a piezo pressure sensor) to initiate the laser illumination prior to the event.

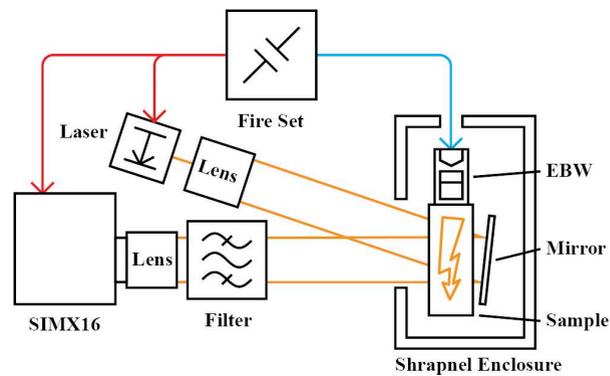

FIG. 6. A schematic of the system used to measure density variations during the DBE.

The choice of a telescope as opposed to an optical relay for imaging of the mirror, was dictated by the short time in which the experimental apparatus could be assembled and operated. Aligning the large number of optical elements in a 45 m optical relay would have left too little time for data acquisition. Future system implementations constructed around an optical relay would enjoy greatly improved light gathering and the attendant SNR improvement. Additionally, placing the objective closer to the object may permit a practical increase in numerical aperture resulting in improved spatial resolution. Two consequences of such a modification would be a reduction in the orthography of the resulting image and depth of field, which may or may not be relevant considerations for specific imaging applications.

## V. CONCLUSION

The ODL system shows promise for accommodating the shutter lag of modern high speed framing cameras. Initial experiments to image dielectric breakdown using a framing camera without an ODL measured high discharge initiation time jitter and involved many frustrating failed imaging attempts. The inclusion of an ODL in the imaging system improved imaging success from 25% to 94%, while also permitting breakdown to be reliably imaged at frame rates 35 times higher than previously achievable in these experiments. With simple modifications and improvements, it is believed that the ODL system will permit reliable imaging of other high-speed phenomena which are either impractical to trigger or suffer from high trigger jitter time. The only requirement being that the phenomenon provides some detectable signal concurrent to the event itself.

## ACKNOWLEDGMENTS

This research was developed with funding from the Defense Advanced Research Projects Agency (DARPA).

## CONFLICT OF INTEREST

The authors have no conflicts to disclose.

## AUTHOR CONTRIBUTIONS

NH was responsible for manuscript preparation, design and implementation of apparatus, and analysis of results. KMS assisted with manuscript preparation, design and implementation of apparatus. AMB and JEK contributed to the design and implementation of the apparatus. AEH, JEB, BLB, SLH, AHM, KS, JPL and BCT contributed to the implementation of the apparatus. JEB, BLB, SLH, AHM, KS, JPL and BCT were responsible for facilitating the experiment and JPL was responsible for camera operation. TWK was responsible for facilitating the experiment, designing and implementation the apparatus and is responsible for the idea behind apparatus' design.

## DATA AVAILABILITY STATEMENT

The data that support the findings of this study are available from the corresponding author upon reasonable request.

## REFERENCES

Akishin, A., Y. Goncharov, Novikov L., Y. Tyutrin, and L. Tseplyayev. 1977. "Discharge Phenomena in Electron-Irradiated Glasses." *Radiation Physics and Chemistry* 319-324.

Cooke, C. 1986. *Space-charge-induced Breakdown in Dielectrics.* Technical Report, Bolling AFB: Air Force Office of Scientific Research.

Cooke, C., E. Williams, and K. Wright. 1982. "Electrical Discharge Propagation in Space-Charged PMMA." *International Conference on Electrical Insulation.* Philadelphia, USA: IEEE. 95-101.

Danikas, M., I. Karafyllidis, A. Thanailakis, and Bruning A. 1996. "Simulation of Electrical Tree Growth in Solid Dielectrics Containing Voids of Arbitrary Shape." *Modelling and Simulation in Materials Science and Engineering* 535-552.

Furuta, J., E. Hiraoka, and S. Okamoto. 1966. "Discharge Figures in Dielectrics by Electron Irradiation." *Journal of Applied Physics* 1873-1878 .

Gross, B. 1958. "Irradiation Effects in Plexiglas." *Journal of Polymer Science* 135-143.

Jenkins, C., R. Ripley, C. Wu, Y. Horie, K. Powers, and W. Wilson. 2013. "Explosively Driven Particle Fields Imaged Using a High Speed Framing Camera and Particle Image Velocimetry." *International Journal of Multiphase Flow* 73-86.

Knaur, J., and P. Budenstein. 1980. "Impulse Breakdown in PMMA under Megavolt, Naosecond Excitation." *IEEE Transactions on Electrical Insulation* 313-321.

Lorenz, A, A K L Dymoke-Bradshaw, and A E Dangor. 1997. "Optical multi-frame system with one gated intensifier as a diagnostic for high-speed photography." *Measurement Science and Technology* 676.

Noskov, M., A. Malinovski, C. Cooke, K. Wright, and A. Schwab. 2002. "Experimental Study and Simulation of Space Charge Stimulated Discharge." *Journal of Applied Physics* 4926-4934.

Settles, G S. 2001. *Schlieren and Shadowgraph Techniques.* 1st. New York: Springer-Verlag.

Smilositz, L., B. Henson, D. Remelius, P. Bowlan, N. Suvorova, J. Allison, D. Cardon, et al. 2020. "Experimental Observations of Exploding Bridgewire Detonator Function." *Journal of Applied Physics.*

Specialised Imaging. 2023. "SIMX ." *full technical specification.* April 27. Accessed April 27, 2023.

Woodworth, J R, J M Lehr, J Elizondo-Decanini, P A Miller, P Wakeland, M Kincy, J Garde, et al. 2004. "Optical and Pressure Diagnostics of 4-MV Water Switches in the Z-20 Test Facility." *IEEE Transactions on Plasma Science* 1778-1789.